\documentclass{ckm}                 

\usepackage{mathptmx}          

\usepackage{epsfig}
\usepackage{rotating}
\usepackage{amsmath}
\usepackage{psfrag}
\allowdisplaybreaks[1]

\usepackage{subfigure}
\renewcommand{\thesubfigure}{(\arabic{subfigure})}

\newcommand{\vcb}{|V_{cb}|}
\newcommand{\vtd}{|V_{td}|}

\newcommand{\vts}{|V_{ts}|}

\def\R1{\varepsilon_1}
\def\E8{\varepsilon_8}

\def\epe{\varepsilon'/\varepsilon}

\newcommand{\mt}{m_{\rm t}}

\newcommand{\mw}{M_{\rm W}}

\newcommand{\gev}{\, {\rm GeV}}
\newcommand{\mev}{\, {\rm MeV}}

\newcommand{\bea}{\begin{eqnarray}}
\newcommand{\eea}{\end{eqnarray}}
\newcommand{\bd}{\begin{displaymath}}
\newcommand{\ed}{\end{displaymath}}

\newcommand{\be}{\begin{equation}}
\newcommand{\ee}{\end{equation}}
\newcommand{\bi}{\begin{itemize}}
\newcommand{\ei}{\end{itemize}}
\newcommand{\ord}{{\cal O}}

\newcommand{\kpnn}{K^+\rightarrow\pi^+\nu\bar\nu}
\newcommand{\kpn}{K^+\rightarrow\pi^+\nu\bar\nu}
\newcommand{\klpn}{K_{\rm L}\rightarrow\pi^0\nu\bar\nu}

\newcommand{\kmm}{K_{\rm L} \to \mu^+ \mu^-}

\confname{Workshop on the CKM Unitarity Triangle, IPPP Durham, April
  2003}

\title{ The Impact of Universal Extra Dimensions on FCNC Processes}

\author{ Andrzej J. Buras\addressmark{a},Anton Poschenrieder\addressmark{a}, 
Michael Spranger \addressmark{a,b} and Andreas Weiler\addressmark{a}}

\address[a]{ Physik Department, Technische Universit{\"a}t M{\"u}nchen, D-85748 Garching, Germany}
\address[b]{ Max-Planck-Institut f{\"u}r Physik - Werner-Heisenberg-Institut, D-80805 M{\"u}nchen, Germany}

\begin{document}

\begin{abstract}
We review the results of two papers on FCNC processes  
in the Appelquist, Cheng and Dobrescu (ACD) model with one universal extra 
dimension. 
\end{abstract}

\maketitle


\section{Introduction}
\setcounter{equation}{0}

Models in more than three spatial dimensions ($D>4$) have been with us for
more than 80 years beginning with the work of Kaluza and Klein, who used this
idea in an attempt to unify gravity and electromagnetism in a five
dimensional model with the extra compact dimension characterized by a radius 
$R$~\cite{Kaluza:tu}. 

While Kaluza, Klein and many authors afterwards considered  very small extra
extra dimensions with the compactification scale $1/R=\ord(M_{\rm Planck})$, 
in recent years there has been an increasing interest in models with
large extra dimensions, in which $1/R=\ord(1~{\rm TeV})$. These models can 
be classified in particular according to the ability of the fields to 
live in extra dimensions. In the so-called ``brane world" models, the SM 
fields live only in the usual four dimensions (``SM on the brane"), while 
gravity lives in the ``bulk", that is in all dimensions~\cite{Antoniadis:1998ig}. In other models
gravity and gauge bosons live in the bulk, while fermions are confined to the
brane. In these models $1/R$ has to be larger than a few TeV in order for the
model to be consistent with the data. Finally, a special role is played by 
models with universal extra dimensions (UED) in which all the SM field 
are allowed
to propagate in all available dimensions. We will concentrate on these
models assuming one extra dimension in what follows.

Above
the compactification scale $1/R$ a given UED model becomes a higher
dimensional field theory whose equivalent description in four dimensions 
consists of the SM fields, the towers of their Kaluza-Klein (KK) partners 
and additional towers of KK modes that do not correspond to any field 
in the SM. Every SM particle has heavy KK partners similar to
the case of the MSSM.

The simplest model of this universal type is the Appelquist, Cheng and 
Dobrescu (ACD) model \cite{Appelquist:2000nn} with one universal extra
dimension. 
In what follows we will briefly describe this model and subsequently report 
on the results of two papers 
\cite{Buras:2002ej,Buras:2003mk} in which we investigated the impact
of the KK modes on FCNC processes in this model. 

\section{The ACD Model}
\setcounter{equation}{0}

The full Lagrangian of this model includes both the bulk and the
boundary Lagrangian. The bulk Lagrangian is determined by the SM
parameters after an appropriate rescaling. The coefficients of
the boundary terms, however, although volume suppressed,
are free parameters and will get renormalized by
bulk interactions. Flavor non-universal boundary terms would lead to large
FCNCs.
In analogy to a common practice in the MSSM where the soft
supersymmetry breaking couplings are chosen to be flavour universal
 we assume negligible
boundary terms 
at the cut-off
scale. With this choice contributions from boundary terms 
are of higher order and we only have to consider the bulk Lagrangian for
the calculation of the impact of the ACD model.

Since all our calculations are cut-off independent (see below) the only additional 
free parameter relative to the SM is the compactification
scale $1/R$. 

Thus all the tree-level masses of the KK particles and their interactions 
among themselves and with the SM particles are described in terms of $1/R$ 
and the parameters of the SM. This economy in new parameters should be 
contrasted with supersymmetric theories and models with an extended Higgs 
sector. All Feynman rules necessary for the evaluation of FCNC processes 
can be found in \cite{Buras:2002ej,Buras:2003mk}.

A very important property of the ACD model is
the conservation of KK parity that implies the absence of tree level 
KK contributions to low energy processes taking place at scales $\mu\ll 1/R$.
In this context the  flavour changing neutral current (FCNC) processes like 
particle-antiparticle mixing, rare K and B decays and radiative decays are 
of particular interest. Since these processes first appear at 
one-loop in the SM and are 
strongly suppressed, the one-loop contributions from the KK modes to them 
could in principle be important.

The effects of the KK modes on various processes of interest have been 
investigated in a number of papers. In
\cite{Appelquist:2000nn,Appelquist:2002wb} their impact  
on the precision electroweak observables assuming 
a light Higgs ($ m_H \le 250~\gev$) and a heavy Higgs led to the lower bound 
$1/R\ge 300\gev$ and $1/R\ge 250\gev$, respectively.
Subsequent analyses of the 
anomalous magnetic moment \cite{Agashe:2001ra} and the  
$Z\to b\bar b$ vertex \cite{Oliver:2002up}
have shown the consistency of the ACD 
model with the data for $1/R\ge 300\gev$. The latter calculation has been 
confirmed in \cite{Buras:2002ej}.
The scale of $1/R$ as low as $300\gev$ would also lead to an exciting
phenomenology in the next generation of colliders and could be of interest 
in connection with dark matter searches. The relevant references are given in
\cite{Buras:2003mk}.

The question then arises whether such low compactification scales are still 
consistent with the data on FCNC processes. This question has been addressed 
in detail in \cite{Buras:2002ej,Buras:2003mk}. Before presenting the results of these
papers let us recall the particle content of the ACD model that 
has been described in detail in  \cite{Buras:2002ej} .

In the effective four dimensional theory, in
addition to the ordinary particles of the SM, denoted as zero $(n=0)$
modes, there are infinite towers of the KK modes $(n\geq 1)$. There is
 one such tower for each SM boson and two for each SM fermion, 
while there also exist physical neutral ($a_{(n)}^0$) and charged 
($a_{(n)}^\pm)$ scalars with $(n\ge1)$ that do not have any zero mode partners.
The masses of the KK particles are universally given by
\be\label{mass}
(m_{(n)}^2)_{\rm KK}=m_0^2+\frac{n^2}{R^2}~.
\ee
Here $m_0$ is the mass of the zero mode, $M_W$, $M_Z$, $m_t$ respectively. 
For $a_{(n)}^0$ and $a_{(n)}^\pm$ this is $M_Z$ and $M_W$, respectively.
In phenomenological applications 
it is more useful to work with the variables $x_t$ and $x_n$ 
defined through
\be\label{xtxn}
x_t=\frac{\mt^2}{\mw^2},\qquad x_n=\frac{m_n^2}{\mw^2},\qquad 
m_n=\frac{n}{R}
\ee
than with the masses in (\ref{mass}).
\section{The ACD Model and FCNC Processes}
\setcounter{equation}{0}
As our analysis of \cite{Buras:2002ej,Buras:2003mk} shows, the ACD
model with one extra dimension
has a number of interesting properties from
the point of view of FCNC processes 
discussed here. These are: 

\begin{itemize}
\item
GIM mechanism \cite{Glashow:gm} that improves significantly the
convergence of  the sum over the KK modes 
corresponding to the top quark, removing simultaneously to an excellent 
accuracy the contributions of the KK modes corresponding to lighter 
quarks and leptons. This feature removes the sensitivity of the calculated
branching ratios to the scale $M_s\gg 1/R$ at which the higher dimensional 
theory becomes non-perturbative and at which the towers of the KK particles 
must be cut off in an appropriate way. This should be contrasted with 
models with fermions localized on the brane, in which the KK parity is not 
conserved and the sum over the KK modes diverges. 
In these models the results are sensitive to $M_s$ and for instance 
in $\Delta M_{s,d}$, the KK effects  are significantly larger
\cite{PaSaOl} than found by us. We expect similar behaviour in other
processes considered below. 
\item
The low energy effective Hamiltonians are governed by local operators 
already present in the SM. As flavour violation and CP violation in 
this model is entirely governed by the CKM matrix, the ACD model belongs 
to the class of the so-called models with minimal flavour violation (MFV) 
as defined in \cite{Buras:2000dm}. This has automatically the
following important consequence for the FCNC processes considered in
\cite{Buras:2002ej,Buras:2003mk}: 
the impact of the KK modes on the processes in question amounts 
only to the modification of the Inami-Lim one-loop functions
\cite{Inami:1980fz}. 
\item
Thus in the case of $\Delta M_{d,s}$ and of the parameter $\varepsilon_K$, 
that are relevant for the standard analysis of the Unitarity Triangle,
these modifications have to be made in the function $S$ \cite{BSS}. In 
the case of the rare K and B decays that are dominated by $Z^0$ penguins 
the functions $X$ and $Y$ \cite{Buchalla:1990qz} receive KK contributions. 
Finally, in the case of 
the decays
$B\to X_s\gamma$, $B\to X_s~ {\rm gluon}$, $B\to X_s\mu\bar\mu$ and 
$K_L\to \pi^0e^+e^-$ and the CP-violating ratio $\epe$ the KK
contributions to new short distance functions have to be
computed. These are  
the functions $D$ (the $\gamma$ penguins), $E$ (gluon penguins), 
$D'$ ($\gamma$-magnetic penguins) and $E'$ (chromomagnetic 
penguins).
\end{itemize}

Thus each function mentioned above, that in the SM depends only on $m_t$, 
becomes now also a function of $1/R$:
\be\label{FACD}
F(x_t,1/R)=F_0(x_t)+\sum_{n=1}^\infty F_n(x_t,x_n), ~~ F=B,C,D,E,D',E',
\ee
with $x_n$ defined in (\ref{xtxn}). 
The functions
$F_0(x_t)$ result from the penguin and box diagrams in the SM and 
the sum represents the KK contributions to these diagrams.

In the phenomenological applications it is 
convenient to work with the gauge invariant functions
\cite{Buchalla:1990qz} 
\begin{equation}\label{xx9} 
X=C+B^{\nu\bar\nu}, \qquad 
Y  =C+B^{\mu\bar\mu}, \qquad
Z =C+\frac{1}{4}D.
\end{equation}

The functions $F(x_t,1/R)$ have been calculated in 
\cite{Buras:2002ej,Buras:2003mk}  with the results given
in table~\ref{inamitab}. 
Our results for the function $S$ have been confirmed in
\cite{Chakraverty:2002qk}. 
For $1/R=300~\gev$, the functions $S$, $X$, $Y$, $Z$ 
are enhanced by $8\%$, $10\%$, $15\%$ and $23\%$ relative to the SM values, 
respectively. The impact of the KK modes on the function $D$ is negligible. 
The function $E$ is moderately enhanced but this enhancement plays only 
a marginal role in the phenomenological applications. The most interesting 
are very strong suppressions of $D'$ and $E'$, that for $1/R=300\gev$ amount 
to $36\%$ and $66\%$ relative to the SM values, respectively. 
However, the  effect of the latter suppressions  is softened in 
the relevant branching ratios through sizable additive QCD corrections.

\begin{table*}[hbt]
\begin{center}
\begin{tabular}{|c||c|c|c|c|c|c|c||c|c|}\hline
 $1/R~[{\rm GeV}]$  & {$S$} & {$X$}& {$Y$} & {$Z$} & {$E$} & {$D'$} & {$E'$} 
& {$C$} & $D$
 \\ \hline
$ 200$ & $ 2.813 $ &  $1.826 $ &  $1.281$ & $0.990$ &$ 0.342$  & $0.113$  & $-0.053$ & $1.099$  & $-0.479$  
\\ \hline
$ 250$ & $ 2.664 $ &  $1.731 $ &  $1.185$ & $0.893$ & $0.327$  & $0.191$ &$ 0.019$ & $1.003$  & $-0.470$
\\ \hline
$300$  & $ 2.582 $ &  $1.674 $ &  $1.128$ & $0.835$ & $0.315$  & $0.242$  &$ 0.065$ & $0.946$  & $-0.468$
\\ \hline
$400$  & $ 2.500 $ &  $1.613 $ &  $1.067$ & $0.771$ & $0.298$  & $0.297$  &$ 0.115 $ & $0.885$  & $-0.469$
\\ \hline
SM     & $2.398$   &  $ 1.526 $ & $0.980$ & $0.679$ & $0.268$  &$ 0.380$ & $ 0.191$ & $0.798$  & $-0.476$
\\ \hline
\end{tabular}
\end{center}
\caption[]{\small \label{inamitab} Values for the functions $S$, $X$, $Y$, 
$Z$, $E$, $D'$, $E'$, $C$ and $D$.
\label{XYZ}}
\end{table*}

\section{The Impact of the KK Modes on Specific Decays}
\setcounter{equation}{0}
\subsection{The Impact on the Unitarity Triangle}
Here the function $S$ plays the crucial role. Consequently the impact 
of the KK modes on the UT is rather small. For $1/R=300\gev$, $\vtd$, 
$\bar\eta$ and $\gamma$ are suppressed by $4\%$, $5\%$ and $5^\circ$, 
respectively. It will be difficult to see these effects in the 
$(\bar\varrho,\bar\eta)$ plane. On the other hand a $4\%$ suppression 
of $\vtd$ means a $8\%$ suppression of the relevant branching ratio for a
rare decay sensitive to $\vtd$ and this effect has to be taken into account. 
Similar comments apply to $\bar\eta$ and $\gamma$. 
Let us also mention that for $1/R=300\gev$, $\Delta M_s$ is enhanced by 
$8\%$ that in view of the sizable uncertainty in $\hat B_{B_s}\sqrt{F_{B_s}}$ 
will also be difficult to see.
\subsection{The Impact on Rare K and B decays}
Here the dominant KK effects enter through the function $C$ or equivalently
$X$ and $Y$, depending on the decay considered. In table~\ref{brtable} 
we show seven branching ratios as functions of $1/R$ for central values of 
all remaining input parameters. The hierarchy of the enhancements of 
branching ratios can easily be explained by inspecting the enhancements of
the functions $X$ and $Y$ that is partially compensated by the suppression 
of $\vtd$ in decays sensitive to this CKM matrix element but fully effective 
in decays governed by $\vts$.

\begin{table*}[hbt]
\vspace{0.4cm}
\begin{center}
\begin{tabular}{|c||c|c|c|c|c|}\hline
{$1/R$ } & {$200\gev$} & {$250\gev$}&  {$300\gev$} 
& {$400\gev$} & SM
 \\ \hline
$Br(\kpn)\times 10^{11}$ & $ 8.70 $ & $8.36$ & $ 8.13$ & $ 7.88$ &  $7.49$ 
\\ \hline
$Br(\klpn)\times 10^{11}$ & $ 3.26 $ & $3.17$ & $ 3.09 $ & $ 2.98$ &  $2.80$ 
\\ \hline
$Br(\kmm)_{\rm SD}\times 10^{9} $ & $ 1.10 $ & $1.00$ &  $ 0.95$ & $ 0.88$ &
$0.79$\\ \hline
$Br(B\to X_s\nu\bar\nu)\times 10^{5}$ & $5.09 $ & $4.56$ & $ 4.26 $ & $ 3.95$ &
$3.53$ \\ \hline
$Br(B\to X_d\nu\bar\nu)\times 10^{6}$ & $1.80 $ & $1.70$ & $ 1.64$ & $ 1.58$ &
$1.47$ \\ \hline
$Br(B_s\to \mu^+\mu^-)\times 10^{9}$ & $6.18 $ & $5.28$ & $ 4.78$ & $4.27$ &
$3.59$ \\ \hline
$Br(B_d\to \mu^+\mu^-)\times 10^{10}$ & $1.56$ & $1.41$ & $ 1.32 $ & $ 1.22$ &
$1.07$ \\ \hline
\end{tabular}
\caption[]{\small Branching ratios for rare decays in the ACD model and the 
SM as discussed in the text.
\label{brtable}}
\end{center}
\end{table*}

For $1/R=300\gev$ the following enhancements relative to the SM predictions 
are seen:
 $\kpn~(9\%)$, $\klpn~(10\%)$, 
$B\to X_{d}\nu\bar\nu~(12\%)$, $B\to X_{s}\nu\bar\nu~(21\%)$, 
$K_L\to\mu\bar\mu~(20\%)$, $B_{d}\to\mu\bar\mu~(23\%)$ and 
$B_{s}\to\mu\bar\mu~(33\%)$.
These results correspond to central values of the 
input parameters.
The uncertainties in these parameters  partly cover the 
differences between the ACD model and the SM model and it is essential 
to reduce these uncertainties considerably if one wants to see the effects 
of the KK modes in the branching ratios in question.

\subsection{An Upper Bound on \bf{$Br(\kpn)$} in the ACD Model}
The enhancement of $Br(\kpn)$ in the ACD model is interesting in view
of the results from the BNL E787
collaboration at Brookhaven \cite{Adler01} that read 
\be\label{kp01}
Br(K^+ \rightarrow \pi^+ \nu \bar{\nu})=
(15.7^{+17.5}_{-8.2})\cdot 10^{-11}
\end{equation}
with the central value  by a factor of 2 above the SM expectation. Even if 
the errors are substantial and this result is compatible with the SM, 
the ACD model with a low compactification scale is  closer to the 
data. 

In \cite{Buchalla:1998ba} an upper bound on $Br(K^+ \rightarrow \pi^+
\nu \bar{\nu})$ 
has been derived within the SM. This bound depends only on $\vcb$, $X$, 
$\xi$ and $\Delta M_d/\Delta M_s$. With the precise value for the angle 
$\beta$ now available this bound can be turned into a useful formula for 
$Br(K^+ \rightarrow \pi^+ \nu \bar{\nu})$ \cite{D'Ambrosio:2001zh}
that expresses 
this branching ratio in terms of theoretically clean observables. 
In the ACD model this formula reads:
\bea \label{AIACD}
Br(K^+ \rightarrow \pi^+ \nu \bar{\nu})\!&\!=\!&\!
\bar\kappa_+\vcb^4 X^2(x_t,1/R)\times\nonumber\\
&&\hspace{-3.2cm}\Bigg[ \sigma
  R^2_t\sin^2\beta+
\frac{1}{\sigma}\left(R_t\cos\beta +
\frac{\lambda^4P_0(X)}{\vcb^2X(x_t,1/R)}\right)^2\Bigg],
\eea
where $\sigma=1/(1-\lambda^2/2)^2$, $\bar\kappa_+=7.5\cdot 10^{-6}$, 
$P_0(X)\approx 0.40$ and

\be\label{Rt}
R_t=0.90~\left[\frac{\xi}{1.24}\right] \sqrt{\frac{18.4/ps}{\Delta M_s}} 
\sqrt{\frac{\Delta M_d}{0.50/ps}},
\quad
\xi = 
\frac{\sqrt{\hat B_{B_s}}F_{B_s} }{ \sqrt{\hat B_{B_d}}F_{B_d}}.
\ee
This formula is theoretically very clean and does not involve 
hadronic uncertainties except for $\xi$  and to a lesser 
extent in $\vcb$. 

In order to find the upper bound on $Br(\kpn)$ in the ACD model we use
$\vcb\le 0.0422$, $P_0(X)<0.47$, $\sin\beta=0.40$ and $\mt<172~\gev$,
where we have set $\sin 2 \beta = 0.734$, its central value as 
$Br(\kpn)$ depends very weakly on it.  
The result of this exercise is shown in table~\ref{Bound}. We give 
there $Br(\kpn)_{\rm max}$  
as a function of $\xi$ and $1/R$ for two different values of $\Delta M_s$.
We observe that for $1/R=250~\gev$ and $\xi=1.30$ the maximal value
for $Br(\kpn)$ in the ACD model is rather close to the central value in
(\ref{kp01}). See \cite{Buras:2002ej} for more details.

\begin{table*}[hbt]
\vspace{0.4cm}
\begin{center}
\begin{tabular}{|c||c|c|c|c|c|}\hline
{$\xi$ } & {$1/R=200~\gev$} & {$1/R=250~\gev$}& {$1/R=300~\gev$} 
& {$1/R=400~\gev$} & SM
 \\ \hline
$1.30$ & $ 13.8^*~(12.3^*) $ & $ 12.7^*~(11.3^*)$ & $ 12.0^*~(10.7)$ & 
$ 11.3^*~(10.1)$ &  $10.8~(9.3)$ \\ \hline
$1.25$ & $ 13.0^*~(11.6) $ & $ 12.0~(10.7) $ & $ 11.4~(10.2) $ & 
$ 10.7~(9.6)$ & $10.3~(8.8)$ \\ \hline
$1.20$ & $ 12.2^*~(10.9) $ & $ 11.3~(10.1) $ & $ 10.7~(9.6) $ & 
$ 10.1~(9.1)$ & $9.7~(8.4)$\\ \hline
$1.15$ & $11.5~(10.3) $ & $10.6~(9.5) $ & $ 10.1~(9.0) $ & $ 9.5~(8.5)$ &
$9.1~(7.9)$ \\ \hline
\end{tabular}
\caption[]{\small Upper bound on $Br(\kpnn)$ in units of $10^{-11}$ for 
different 
values of $\xi$, $1/R$ and $\Delta M_s=18/{\rm ps}~(21/{\rm ps})$. The 
stars indicate the results corresponding to 
$\sqrt{\hat B_{B_d}} F_{B_d}\le 190\mev$.
\label{Bound}}
\end{center}
\end{table*}

\subsection{The Impact on \bf{$B\to X_s\gamma$} and
   \bf{$B\to X_s~{\rm gluon}$}}

The inclusive $B\to X_s\gamma$ decay has been the subject of very intensive 
theoretical and experimental studies during the last 15 years. On the 
experimental side the world average resulting from the data by CLEO, ALEPH, 
BaBar and Belle reads \cite{Battaglia:2003in}
\be\label{bsgexp}
Br(B\to X_s\gamma)_{E_\gamma> 1.6{\rm GeV}}=
 (3.28^{+0.41}_{-0.36})\cdot 10^{-4}~.
\ee
It agrees well with the SM result \cite{Gambino:2001ew,Buras:2002tp}
\be\label{bsgth}
Br(B\to X_s\gamma)^{\rm {SM}}_{E_\gamma> 1.6{\rm GeV}}
= (3.57\pm 0.30)\cdot 10^{-4}~.
\ee
The most recent reviews summarizing the theoretical status can be found 
in \cite{Buras:2002er,ALIMIS,Hurth:2003vb}.

Due to strong suppressions of the functions $D'$ and $E'$ by the KK modes, 
the $B\to X_s\gamma$ and $B\to X_s~{\rm gluon}$ decays are considerably 
suppressed compared to SM estimates. For $1/R=300\gev$, 
$Br(B\to X_s\gamma)$ is suppressed by $20\%$, while $Br(B\to X_s~{\rm gluon})$
even by $40\%$. The phenomenological relevance of the latter suppression is 
unclear at present as $Br(B\to X_s~{\rm gluon})$ suffers from large 
theoretical uncertainties and its extraction from experiment is very
difficult if not impossible.

\begin{figure}[]
\renewcommand{\thesubfigure}{\space(\alph{subfigure})}
\centering
\psfragscanon

  \psfrag{bsgammabsgammabsg}{ $Br(B\rightarrow X_s \gamma )\times 10^{4}$}
  \psfrag{rinvrinv}[][]{ \shortstack{\\  $R^{-1}$ [GeV] }}
  \label{bsg.eps}
        \resizebox{.36\paperwidth}{!}{\includegraphics[]{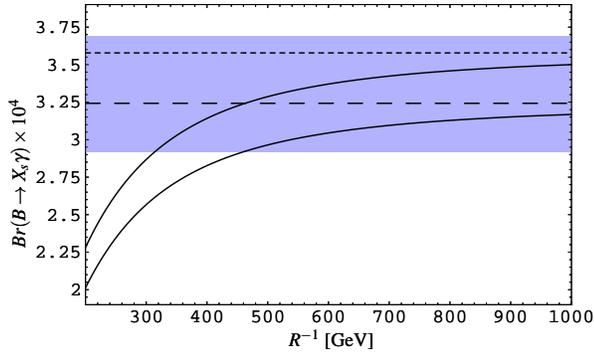}}
    \caption[]{\small\label{bsgplot} The branching ratio for $B\to
      X_s\gamma$  and $E_\gamma > 1.6$ GeV as a function of $1/R$. 
See text for
the meaning of various curves.}
  \end{figure}

In fig.~\ref{bsgplot} we compare $Br(B\to X_s\gamma)$ 
in the ACD model with the experimental data and with the expectations 
of the SM. The shaded region represents the data in (\ref{bsgexp}) and the 
upper (lower) dashed horizontal line are the central values in the SM 
for $m_c/m_b=0.22~(m_c/m_b=0.29)$. The solid lines represent the 
corresponding central values in the ACD model. The theoretical errors, 
not shown in the plot, are for all curves roughly $\pm 10\%$

We observe that in view 
of the sizable experimental error and considerable parametric uncertainties in 
the theoretical prediction, the strong suppression of $Br(B\to X_s\gamma)$ 
by the KK modes does not yet provide a powerful lower bound on $1/R$ and the
values $1/R\ge 250\gev$ are fully consistent with the experimental result. 
It should also be emphasized that $Br(B\to X_s\gamma)$ depends
sensitively on the ratio $m_c/m_b$ and the lower bound on $1/R$ is shifted above 
$400\gev$ for $m_c/m_b=0.29$ if other uncertainties are neglected. 
In order to reduce the dependence on $m_c/m_b$ a NNLO calculation 
is required. Once it is completed and the experimental uncertainties reduced, 
$Br(B\to X_s\gamma)$ may provide a very powerful bound on $1/R$ that is 
substantially stronger than the bounds obtained from the electroweak precision 
data.

The suppression of $Br(B\to X_s\gamma)$ in the ACD model has already been
found in \cite{Agashe:2001xt}. Our result presented above
is consistent with the one obtained by these authors but differs in details 
as only the dominant diagrams have been taken into account in the latter paper 
and the analysis was performed in the LO approximation.

\subsection{The Impact on \bf{$B\to X_s\mu^+\mu^-$} and
   \bf{$A_{FB}(\hat s)$}}
The inclusive $B\to X_s\mu^+\mu^-$ decay has been the subject of very 
intensive theoretical and experimental studies during the last 15 years. On 
the 
experimental side only the BELLE collaboration reported the observation of this 
decay with \cite{Kaneko:2002mr}
\be\label{bsxmmexp}
Br(B\to X_s\mu^+\mu^-)= (7.9\pm 2.1^{+2.0}_{-1.5})\cdot 10^{-6}~.
\ee 
For the decay to be
dominated by perturbative contributions one has to remove $\bar c c$
resonances by appropriate kinematic cuts in the dilepton mass
spectrum.
The SM expectation \cite{Ali:2002jg} for the low dilepton mass window
is given by 
\be\label{bsxmmth}
\tilde Br(B\to X_s\mu^+\mu^-)_{\rm SM}= (2.75\pm 0.45)\cdot 10^{-6}~
\ee
where the dilepton mass spectrum has been integrated between the limits:
\be\label{intlimits}
\left(\frac{2 m_\mu}{m_b}\right)^2\le\hat s\le
\left(\frac{M_{J/\psi}-0.35\gev}{m_b}\right)^2
\ee
where $\hat s=(p_++p_-)^2/m_b^2$.

This cannot be directly compared to the
experimental result in (\ref{bsxmmexp}) that is supposed to include the
contributions from the full dilepton mass spectrum. 
Fortunately future experimental
analyses should give the results corresponding to the low dilepton 
mass window so that a direct comparison between the experiment and the 
theory will be possible.
The most recent reviews summarizing the theoretical status can be found 
in \cite{Hurth:2003vb,Ali:2002jg}. 

\begin{figure}[hbt]
\renewcommand{\thesubfigure}{\space(\alph{subfigure})} 
  \centering 
\psfragscanon
  \psfrag{brsllbrsllbrsll}{  $\tilde Br(B\rightarrow X_s \mu^+ \mu^- )\times 10^{5}$}
  \psfrag{rinvrinv}[][]{ \shortstack{\\ $R^{-1}$ [GeV] }}
    \resizebox{.36\paperwidth}{!}{
     \includegraphics[]{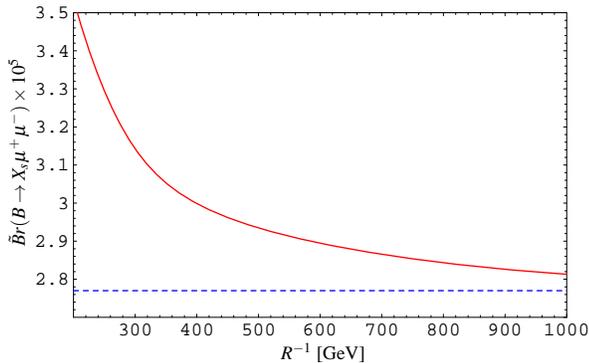}}
    
  \caption[]{\small\label{brsllRinv} $\tilde Br(B\rightarrow X_s \mu^+
  \mu^- )$  in the SM (dashed line) and in the ACD model. The
  integration limits are discussed in the text.}
\end{figure}

In fig.~\ref{brsllRinv} we show the branching ratio 
$\tilde Br(B\to X_s\mu^+\mu^-)$ as a function
of $1/R$ that corresponds to the SM result of (\ref{bsxmmth}). 
The observed enhancement is 
mainly due to the function $Y$ that enters the Wilson coefficient of the
operator $(\bar s b)_{V-A}(\bar\mu \mu)_A$. The Wilson coefficient 
of $(\bar s b)_{V-A}(\bar\mu \mu)_V$, traditionally denoted by $C_9$, is 
essentially unaffected by the KK contributions. 

Of particular interest is the Forward-Backward asymmetry $A_{\rm FB}(\hat s)$
in $B\to X_s\mu^+\mu^-$ that 
similarly to the case of exclusive 
decays \cite{Burdman:1998mk} vanishes at 
a particular value $\hat s=\hat s_0$. 
The fact that $A_{\rm FB}(\hat s)$ and the value of $\hat s_0$ 
being sensitive to short distance physics are in addition 
subject to only very small non-perturbative uncertainties makes them 
particularly useful quantities to test physics beyond the SM. 

\begin{figure}[hbt]
\renewcommand{\thesubfigure}{\space(\alph{subfigure})} 
  \centering 
  \subfigure[]{\psfragscanon
  \psfrag{nfbnfb}{ $\hat{A}_{FB}$}
  \psfrag{hats}[][]{ \shortstack{\\ $\hat{s} $ }}
      \resizebox{.36\paperwidth}{!}{\includegraphics[]{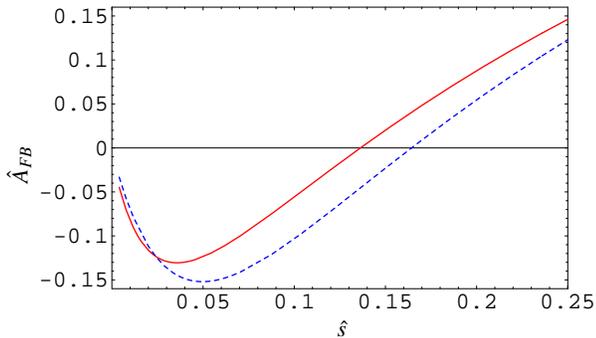}}
    }
  \subfigure[]{\psfragscanon
  \psfrag{zafb}{ $\hat{s}_{0}$}
  \psfrag{rinvrinv}[][]{ \shortstack{\\  $R^{-1}$ [GeV]  }}
    \resizebox{.36\paperwidth}{!}{ \includegraphics[]{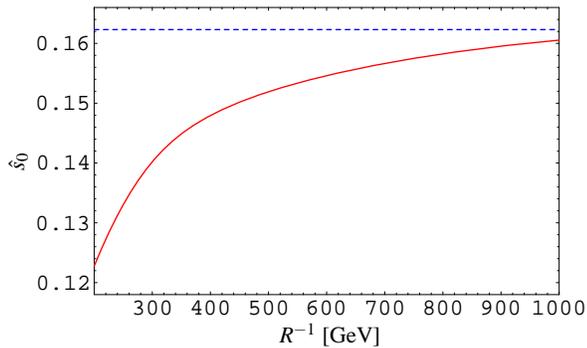}}
    }
  \caption[]{\small\label{normalizedfb} (a) Normalized Forward-Backward
    asymmetry in the SM (dashed line) and ACD for $R^{-1}=250$ GeV. 
(b) Zero of the forward backward
    asymmetry $A_{FB}$ in the SM (dashed line) and the ACD model.}
\end{figure}

The calculations for $A_{\rm FB}(\hat s)$ and of $\hat s_0$ have
recently been 
done including NNLO corrections \cite{NNLO1,NNLO2} that turn out to 
be significant. 
In particular they shift the NLO value of $\hat s_0$ from $0.142$ to
$0.162$ at NNLO. 
In fig.~\ref{normalizedfb} (a)  we show the normalized
Forward-Backward asymmetry that we obtained by means of the formulae and 
the computer program of 
\cite{Ali:2002jg,NNLO1} modified by the 
KK contributions calculated in~\cite{Buras:2003mk}. The dependence of 
$\hat s_0$ on $1/R$  
is shown in fig.~\ref{normalizedfb} (b).

We observe that the value of $\hat s_0$ is considerably reduced relative 
to the SM result obtained by including NNLO corrections 
\cite{Ali:2002jg,NNLO1,NNLO2}. This decrease is related to the decrease 
of $Br(B\to X_s\gamma)$ as discussed below. 
For 
$1/R=300\gev$ we find the value for $\hat s_0$ that is  very close to 
the NLO prediction of the 
SM. This result demonstrates very clearly the importance of the
calculations of the higher 
order QCD corrections, in particular in quantities like $\hat s_0$
that are theoretically clean. We expect that the results in
figs.~\ref{normalizedfb} (a) and (b) 
will play an important role in the tests of the ACD model in the future.

\begin{figure}[hbt]

  \centering 
 \psfragscanon
  \psfrag{bsgammabsgammabsgamma}{ \shortstack{\\ \\ $(Br(B\to
 X_s\gamma)\times 10^4)^\frac12$ ${}$}}
  \psfrag{hats0}[][]{ \shortstack{\\ $\hat{s}_0 $ }}
      \resizebox{.36\paperwidth}{!}{\includegraphics[]{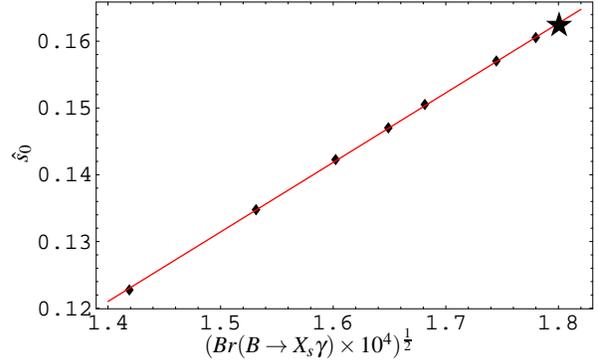}}
    
  \caption[]{\small\label{corrplot} Correlation between
    $\sqrt{Br(B\to X_s\gamma)}$  and $\hat s_0$. The straight line is
    a least square fit to a linear function. The dots are the results in the ACD
    model
    for $1/R = 200,250,300,350,400,600$ and $1000$ GeV  and the star
    denotes the SM value.
}
\end{figure}

In MFV models there exist a number of correlations between different 
measurable 
quantities that do not depend on specific parameters of a given model
\cite{Buras:2000dm,REL}. In \cite{Buras:2003mk}
a correlation between $\hat s_0$ and 
$Br(B\to X_s\gamma)$ has been pointed out. It is present in the ACD model 
and in a large 
class of supersymmetric models discussed for instance in
\cite{Ali:2002jg}.  
We show this correlation in fig.~\ref{corrplot}. We refer to
\cite{Buras:2003mk} for further details.

\subsection{The Impact on $\boldmath{K_L\to\pi^0 e^+ e^-}$ and 
$\boldmath{\epe}$}
The impact of the KK modes on the rare decay $K_L\to\pi^0 e^+ e^-$ 
is at most $10\%$ but it is substantially larger on $\epe$.
The most recent discussion on $\epe$ can be found in \cite{BJ03}.
As the $Z^0$ penguins are enhanced in the ACD model, the ratio 
$\epe$ is suppressed relative to the SM expectations with the size of the 
suppression depending sensitively on the hadronic matrix elements. 
In view of this no useful bound on $1/R$ can be obtained from $\epe$ 
at present. 

\section{Concluding Remarks}

Our analysis of the ACD model shows that all the present data on FCNC 
processes are consistent with $1/R$ as low as $250\gev$, implying that 
the KK particles could in principle be found already at the Tevatron.
Possibly, the most interesting results of our analysis is the 
enhancement of $Br(\kpn)$, the sizable downward 
shift of  the zero ($\hat s_0$) in the 
$A_{\rm FB}$ asymmetry and the suppression of $Br(B\to X_s\gamma)$.

The nice feature of this extension of the SM is the presence of only one 
additional parameter, the compactification scale. This feature allows a
unique determination of various
enhancements and suppressions relative to the SM expectations. 
We find
\begin{itemize}
\item
Enhancements: $K_L\to \pi^0e^+e^-$,  $\Delta M_s$,
 $\kpn$, $\klpn$, $B\to X_{d}\nu\bar\nu$, $B\to X_{s}\nu\bar\nu$, 
$K_L\to\mu^+\mu^-$, $B_{d}\to\mu^+\mu^-$, $B\to X_s\mu^+\mu^-$ and 
$B_{s}\to\mu^+\mu^-$.
\item
Suppressions: 
$B\to X_s\gamma$, $B\to X_s~{\rm gluon}$, the value of $\hat s_0$ in the 
forward-backward asymmetry and $\epe$.
\end{itemize}

We would like to emphasize that violation of this pattern by the future 
data will
exclude the ACD model. For instance the measurement of $\hat s_0$ that is 
higher than the SM estimate would automatically exclude this model as 
there is no compactification scale for which this could be satisfied.
Whether these enhancements and suppressions are required by the data or 
whether they exclude the ACD model with a low compactification scale, 
will depend 
on the precision of the forthcoming experiments and the efforts to decrease 
the theoretical uncertainties.

\section*{Acknowledgements}
This
research was partially supported by the German `Bundesministerium f\"ur 
Bildung und Forschung' under contract 05HT1WOA3 and by the 
`Deutsche Forschungsgemeinschaft' (DFG) under contract Bu.706/1-2.


\newcommand{\np}[3]{Nucl.~Phys. {\bf B#1} (#2) #3}
\newcommand{\pl}[3]{Phys.~Lett. {\bf B#1} (#2) #3}
\newcommand{\pr}[3]{Phys.~Rev.  {\bf D#1} (#2) #3}
\newcommand{\prl}[3]{Phys.~Rev. Lett. {\bf #1} (#2) #3}
\newcommand{\prp}[3]{Phys.~Rept. {\bf #1} (#2) #3}
\newcommand{\zpc}[3]{Z.~Phys. {\bf C#1} (#2) #3}
\newcommand{\hep}[2]{[arXiv:hep-#1/#2]}

\renewcommand{\baselinestretch}{0.95}

\vfill\eject

\end{document}